\documentclass[11pt]{article}
\usepackage{epsfig}
\usepackage{amsmath,latexsym}

\def\sect#1{\section{#1}\setcounter{equation}{0}}

\begin{document}
\title{Numerical Implementation of the Multisymplectic Preissman
Scheme and Its Equivalent Schemes} \footnotetext {Supported by the
National Natural Sciences Foundation of China (No. 49825109) the
CAS Key Innovation Direction Project (No. KZCX2-208) and National
Key Development Planning Project for the Basic Research (No.
199032081)
 }
\author{ Yushun Wang\quad Bin Wang\\
Lasg, Institute of Atmospheric Physics,\\Chinese Academy of
Sciences, Beijing 100029,China\\
Mengzhao Qin \\
Lsec, School of mathematics and system science,\\ Chinese Academy
of Science, 100080, China
 }
\date{}
\maketitle
\begin{abstract}
We analyze the multisymplectic Preissman scheme for the KdV
equation with the periodic boundary condition and show that the
unconvergence of the widely-used iterative methods to solve the
resulting nonlinear algebra system of the Preissman scheme is due
to the introduced potential function. A artificial numerical
condition is added to the periodic boundary condition. The added
boundary condition makes the numerical implementation of the
multisymplectic Preissman scheme practical and is proved not to
change the numerical solutions  of the KdV equation. Based on our
analysis, we derive some new schemes which are not restricted by
the artificial boundary condition and more efficient than the
Preissman scheme because of less computing cost and less computer
storages. By eliminating the auxiliary variables, we also derive
two schemes for the KdV equation, one is a 12-point scheme and the
other is an 8-point scheme.  As the byproducts, we present two new
explicit schemes which are not multisymplectic but still have
remarkable numerical stable property. Numerical experiments on
soliton collisions are also provided to confirm our conclusion and
to show the benefits of the multisymplectic schemes with
comparison of the spectral method and Zabusky-Kruskal scheme.
\end{abstract}

\sect{Introduction} The Korteweg-de Veris equation has been used
to describe various  phenomena such as acoustic waves in an
anharmonic crystal, waves in bubble-liquid mixtures,
magnetohydrodynamic waves in warm plasma, and ion acoustic waves.
This equation has two fascinating
 and significant features. One is the existence of permanent wave solutions, including
solitary wave solutions, and the other is the recurrence of the
initial state of the wave form. In 1965 Zabusky and Kruskal
\cite{zk} used a finite difference method, i.e. the famous
Zabusky-Kruskal scheme, to show the existence of solitons which
propagate with their own velocities, exerting essentially no
influence on each other. They also discussed the recurrence of an
initial state and guessed that the KdV equation led to the
recurrence. Since then, various  methods including the finite
difference method, the Fourier expansion method \cite{fouri} and
the finite element method \cite{rw} have been proposed to solve
the KdV equation. Unfortunately, the difference solutions often
exhibit nonlinear instabilities when a long time integration is
performed.
 In the 1990s, the symplectic schemes were introduced and
systematically developed for the Hamiltonian systems within the
framework of symplectic geometry \cite{hai}-\cite{fq}. Numerical
results show that symplectic schemes have superior performance,
especially in long time simulations. The symplectic schemes can be
applied to the KdV equation which may be transformed into the form
of Hamiltonian system.

Recently,  J. E. Marsden etc. \cite{mars} and T. J. Bridges etc.
\cite{bs} proposed the concept of multisymplectic PDEs and
multisymplectic schemes
 which can be viewed as the generalization
of symplectic schemes. Many soliton equations such as the KdV
equation, the Kadomtsev-Petviashvili equation, the
Zabolotskaya-Khokhlov equation and the sine-Gordon equation can be
reformulated into the multisymplectic PDEs and can be solved
numerically by the multisymplectic schemes. The simplest and basic
multisymplectic scheme is the Preissman scheme, which has been hot
in the last two years \cite{zhao}-\cite{my}. However, sometimes
the direct numerical implementation of the Preissman scheme for
the multisymplectic equation has puzzled researchers all along.
When it is applied to solve the periodic boundary problem of the
soliton equations with degenerate lagrangian like the KdV, K-P
equation and the water waves equation, the general widely-used
iterative method refereed as the simple iterative method
\cite{fouri} is not convergent, so are the other iterative method
such as Newton method and conjugate gradient method. Why?

To settle the problem, S. Reich \cite{sr} eliminated the auxiliary
variables to get a equivalent scheme for the KdV equation, but he
did not consider the influence of the boundary condition.

In the present paper, taking the KdV equation as an example, we
analyze the practical computation of the Preissman scheme and find
that the unconvergence is due to the indeterminacy of the
potential function. We  add a condition on the potential function
to fix it up and prove that the added condition will not change
the numerical solutions of the KdV equation. This condition can be
stated as a  restriction on  numerical periodic  boundary
condition. Based on our analysis, we present some new
multisymplectic implicit schemes that are equivalent to, but more
efficient than the Preissman scheme because of less computing cost
and less computer storages . By converting the implicit term in
the multisymplectic schemes to an explicit one, we obtain two
 stable, efficient, explicit schemes for the KdV equation. Of
 cause, they are not multisymplectic any more.
 An elementary but useful method to eliminate the auxiliary
 variables of the multisymplectic schemes is also presented and
 two new  multisymplectic schemes  for the KdV equation are
 derived. One is a 12-point scheme, the other is an 8-point scheme.

The main purpose of this paper is to develop a method to analyze
the multisymplectic scheme for the Hamiltonian PDEs and to show
how to choose the proper numerical boundary condition for the
multisymplectic schemes and how to derive the new schemes for the
PDEs.  The method presented in the paper can be applied to the
Preissman scheme for other PDEs and  to other multisymplectic
schemes. Another aim is to compare the performance of the
multisymplectic schemes with other kind numerical method to see if
the multisymplectic schemes  benefit  the finite difference
approximations of the PDEs. Thus a series of numerical experiments
on soliton collisions are presented. Compared with the
Zabusky-Kruskal scheme and the spectral method, the
multisymplectic schemes are shown to have superior stability,
excellent ability to preserve the conservation laws and
remarkable capacity of long time computing.

This paper is organized as follows.
In section 2 we take a brief review of multisymplectic
structure of the KdV equation and the multisymplectic Preissman scheme.
 The Preissman scheme is analyzed in section 3, where we present
an artificial numerical boundary condition for the Preissman
scheme and verify its rationality. In section 4, some new
multisymplectic schemes for the KdV equation are derived. Section
5 is  for numerical experiments and we finish the paper with
concluding remarks in section 6.

\sect{Multisymplectic structure of the KdV equation and the Preissman scheme}
 The general form of the KdV equation with the initial value
and the periodic boundary condition is
 \begin{equation}\label{2.1}
\frac{\partial u}{\partial t}+\eta u\frac{\partial u}{\partial
x}+\delta^2 \frac{\partial^3 u}{\partial x^3}=0, \;t>0,
\end{equation}
$$u(t=0, x)=u_0(x),\quad u(t,x+a)=u(t,x+b),$$
where $\eta$ and $\delta$ are two real numbers.

Introducing the potential $\phi_x=u$, momenta $v=\delta u_x$ and
variable $ w={\frac{1}{2}} \phi_t+\delta v_x+V^{'}(u)$, $V(u)=\eta
u^3/6$, the KdV equation (\ref{2.1}) can be rewritten as the
following Hamiltonian PDEs.
\begin{equation}\label{2.2}
  M{\bf z}_{t}+K{\bf z}_{x}=\bigtriangledown_{\bf z}S({\bf z}),
\end{equation}
where
$$M=\left[\begin{array}{cccc}
                  0&{1}\over{2}&0&0\\
                -{{1}\over{2}}&0&0&0\\
                0&0&0&0\\
                 0&0&0&0
                  \end{array}
                  \right],
\quad K=\left[\begin{array}{cccc}
               0&0&0&1\\
               0&0&-\delta&0\\
               0&\delta&0&0 \\
               -1&0&0&0
               \end{array}
               \right],
\quad {\bf z}=\left[\begin{array}{c}
               \phi\\
                u\\
               v\\
               w
               \end{array}
               \right]$$
and
$S({\bf z})=\frac{1}{2}v^2-uw+V(u).$

 Each of the two skew-symmetric
matrices $M$ and $K$ can be identified with a closed two forms.
\begin{equation*}
\omega ^1({\bf u},{\bf v})=<M{\bf u},{\bf  v}>,\qquad
\omega ^2({\bf u},{\bf v})=<K{\bf u},{\bf  v}>,\qquad\qquad
\end{equation*}
where ${\bf u}$, ${\bf v}$ are any vectors on ${\bf R}^4$ and $<\cdot\;, \cdot>$ is
the standard Euclidean inner product on  ${\bf R}^4$.

Both forms $\omega ^i$ $i=1,2$ are closed and therefore
pre-symplectic
 on  ${\bf R}^4$,
and on  subspaces where they are non-degenerated, they are
symplectic forms. In other words, $({\bf R}^2,\omega^1)$, and
$({\bf R}^4,\omega^2)$ are two distinct symplectic manifolds.
Moreover, each two forms is associated with a different direction.
$\omega ^1$ is associated with time and $\omega ^2$ is associated
with space. In this sense the first order PDEs (\ref{2.2}) is
called multisymplectic PDEs or Hamiltonian PDEs. The KdV  equation
is completely characterized by the function $S({\bf z})$, and the
two skew-symmetric operators $M$ and  $K$. They are all defined on
a finite dimensional space.

The multisymplectic Hamiltonian equation (\ref{2.2}) satisfies the
important multisymplectic conservation law
\begin{equation}\label{2.3}
\partial_t[d{\bf z}\wedge Md{\bf z}]+\partial_x[d{\bf z}\wedge Kd{\bf z}]=0,
\end{equation}
which, for the KdV equation (\ref{2.1}), is equivalent to
\begin{equation}\label{2.4}
 \partial_t[d\phi\wedge du]+2\partial_x[d\phi\wedge dw+\delta dv\wedge
 du]=0,
\end{equation}
where $\wedge$ is the standard exterior product operator of the
differential forms.

The conservation law (\ref{2.4}) is a strictly local conservation
concept that does not depend on a specific boundary condition.
That is to say, in the arbitrary domain of the space-time plane,
changes in the wedge product $ d\phi\wedge du$ in time are exactly
compensated for by changes in the wedge product $-2(d\phi\wedge
dw+\delta dv\wedge du)$ in space.

Bridges and Reich \cite{bs} showed that the Preissman scheme for
(\ref{2.2}) is a multisymplectic scheme which preserves the
discrete form of (\ref{2.3}). The  Preissman scheme for
(\ref{2.2}) is
\begin{equation}\label{2.5}
\frac{1}{\tau}M({\bf z}_{n+\frac{1}{2}}^{m+1}-{\bf z}_{n+\frac{1}{2}}^{m})+
\frac{1}{h}K({\bf z}_{n+1}^{m+\frac{1}{2}}-{\bf z}_{n}^{m+\frac{1}{2}})
=\bigtriangledown_{\bf z}S({\bf z}_{n+\frac{1}{2}}^{m+\frac{1}{2}}),
\end{equation}
where $\tau$ is the time step, $h$ is the space step, $x_n,
\;n=1,2,\cdots, N$; $t_m,\;m=1,2,\cdots$ is the regular grids of
the integral domain,  ${\bf z}^m_n$ is an approximation to ${\bf
z}(x_n, t_m)$, ${\bf z}_{n+\frac{1}{2}}^{m}=\frac{1}{2} ({\bf
z}_{n+1}^{m}+{\bf z}_{n}^{m})$, ${\bf
}z_{n}^{m+\frac{1}{2}}=\frac{1}{2} ({\bf z}_{n}^{m+1}+{\bf
z}_{n}^m)$, ${\bf z}_{n+\frac{1}{2}}^{m+\frac{1}{2}}=\frac{1}{4}
({\bf z}_{n}^{m}+{\bf z}_{n+1}^{m}+{\bf z}_{n}^{m+1}+{\bf
z}_{n+1}^{m+1})$, ${\bf z}=(\phi,u,v,w)^T$,
 and the corresponding discretized multisymplectic conservation law  is
\begin{align*}
&\frac{d\phi^{m+1}_{n+\frac{1}{2}}\wedge du^{m+1}_{n+\frac{1}{2}}-
d\phi^m_{n+\frac{1}{2}}\wedge du^m_{n+\frac{1}{2}}}{\tau}=\\
&-2\frac{d\phi^{m+\frac{1}{2}}_{n+1}\wedge
dw^{m+\frac{1}{2}}_{n+1}+\delta dv^{m+\frac{1}{2}}_{n+1}\wedge
du^{m+\frac{1}{2}}_{n+1}-
       d\phi^{m+\frac{1}{2}}_{n}\wedge
dw^{m+\frac{1}{2}}_{n}-\delta dv^{m+\frac{1}{2}}_{n}\wedge
du^{m+\frac{1}{2}}_{n}}{h}.
\end{align*}

\sect{Analysis of the Preissman Scheme } The Preissman scheme
(\ref{2.5}) is an implicit scheme that involves solving a
nonlinear equations for ${\bf z}^{m+1}$ at each time step. The
widely-used iterative method of this nonlinear equations is as
follows. (The analysis and the comparison between  this iterative
technique with other iterative methods such as Newton's Method can
be found in \cite{fouri}.)
\begin{align}
&\frac{1}{2}(u^{j+1}_{i}+u^{j+1}_{i+1})-r
(w^{j+1}_{i}-w^{j+1}_{i+1})=\frac{1}{2}(u^{j}_{i}+u^{j}_{i+1})+r(w^{j}_{i}-w^{j}_{i+1}),
\notag\\
&\frac{\tau}{2}(w^{j+1}_{i}+w^{j+1}_{i+1})-\frac{1}{2}(\phi^{j+1}_{i}+
\phi^{j+1}_{i+1})+\delta r(v^{j+1}_{i}-v^{j+1}_{i+1})=\notag\\
&\quad
-\frac{\tau}{2}(w^{j}_{i}+w^{j}_{i+1})-\frac{1}{2}(\phi^{j}_{i}+
\phi^{j}_{i+1})-\delta r(v^{j}_{i}-v^{j}_{i+1})+2\tau
V^{'}(\bar{u}_i^j),\label{3.1} \\   &\delta
(-u^{j+1}_{i}+u^{j+1}_{i+1})-\frac{h}{2}(v^{j+1}_{i}+v^{j+1}_{i+1})
=\delta (u^{j}_{i}-u^{j}_{i+1})+\frac{h}{2}(v^{j}_{i}+v^{j}_{i+1}),\notag\\
&\frac{h}{2} (u^{j+1}_{i}+u^{j+1}_{i+1})+(\phi^{j+1}_{i}-\phi^{j+1}_{i+1})
=-\frac{h}{2} (u^{j}_{i}+u^{j}_{i+1})-(\phi^{j}_{i}-\phi^{j}_{i+1}),\notag
\end{align}
where $i=1, 2, \cdots, n$,
$\bar{u}_i^j=\frac{1}{4}(u^{j+1}_{i}+u^{j+1}_{i+1}+
u^{j}_{i}+u^{j}_{i+1})$,  $r=\frac{\tau}{h}$ is the ratio between
temporal and spatial steps.

Here we discuss the numerical boundary conditions.

By $ \phi_x=u$,$$\phi(b,t)=\phi(a,t)+\int^b_a u(x,t) dx.$$ Set  $
c =\int^b_a u(x,t) dx$, then
\begin{align*}
\frac{d}{dt} c&=\int^b_a u_t dx\\
&=-\int^b_a (cuu_x+\delta^2u_{xxx}) dx\\
&=-\int^b_a (\frac{c}{2}u^2+\delta^2 u_{xx})_x dx\\
&=0.
\end{align*}

We obtain $c=\int^b_a u_0(x) dx=a\; constant. $

Thus the periodic numerical boundary conditions are
\begin{equation}\label{3.2}
u_1^{j+1}=u_{n+1}^{j+1},\;\; v_1^{j+1}=v_{n+1}^{j+1}, \;\;
\phi_1^{j+1}=\phi_{n+1}^{j+1}+c, \;\; w_1^{j+1}=w_{n+1}^{j+1}.
\end{equation}

Let $$A=\left[
\begin{array}{cccccc}
1&1&0&\cdots&0&0\\
0&1&1&\cdots&0&0\\
\cdot &\cdot &\cdot &\cdots &\cdot &\cdot\\
0&0&0&\cdots&1&1\\
1&0&0&\cdots&0&1
\end{array}
\right]_{n\times n},\;B=\left[
\begin{array}{cccccc}
-1&1&0&\cdots&0&0\\
0&-1&1&\cdots&0&0\\
\cdot &\cdot &\cdot &\cdots &\cdot &\cdot\\
0&0&0&\cdots&-1&1\\
1&0&0&\cdots&0&-1
\end{array}
\right]_{n\times n},$$
${\bf u}^j=(u_1^j, u_2^j,\cdots,
u_n^j)^{T},\; {\bf v}^j=(u_1^j, v_2^j,\cdots, v_n^j)^{T},\; {\bf
w}^j=(u_1^j, w_2^j,\cdots, w_n^j)^{T}, \Phi^j=(\phi_1^j,
\phi_2^j,\cdots, \phi_n^j)^{T},\; {\bf V}({\bf u}^j,{\bf
u}^{j+1})=(V_1, V_2,\cdots, V_n)^{T},
 V_i=V^{'}(\frac{1}{4}(u_i^j+u_i^{j+1}+u_{i+1}^j+u_{i+1}^{j+1})),
 \; i=1,2,\cdots,n-1,\;
V_n=V^{'}(\frac{1}{4}(u_n^j+u_n^{j+1}+u_{1}^j+u_{1}^{j+1})),\;
{\bf c}=(0,\cdots,0,2c)^{T}$ an $n$ dimensional vector,  then
(\ref{3.1}) and (\ref{3.2}) can be rewritten as the following
vector form.
\begin{align}\label{3.3}
\begin{split}
&\frac{h}{2}A{\bf u}^{j+1}-B\Phi^{j+1}=-\frac{h}{2}A{\bf u}^{j}+
B\Phi^{j}+{\bf c},\\
&\delta B{\bf u}^{j+1}-\frac{h}{2}A{\bf v}^{j+1}=-\delta B{\bf u}^j+\frac{h}{2}
A{\bf v}^j,\\
&-\delta r B{\bf v}^{j+1}+\frac{\tau}{2}A{\bf w}^{j+1}-\frac{1}{2}A\Phi^{j+1}
=\delta r B{\bf v}^{j}-\\
&\qquad\qquad\frac{\tau}{2}A{\bf w}^{j}-\frac{1}{2}A\Phi^j+
2\tau{\bf V}({\bf u}^j,{\bf u}^{j+1}),\\
&\frac{1}{2}A{\bf u}^{j+1}+rB{\bf w}^{j+1}=\frac{1}{2}A{\bf u}^j-rB{\bf w}^j.
\end{split}
\end{align}

Also let ${\bf X}=({\bf u}^{j+1}, {\bf v}^{j+1}, {\bf w}^{j+1}, {\bf\phi} ^{j+1})^{T}$,
we get the iterative form of the nonlinear equations
\begin{equation}\label{3.4}
D{\bf X}^{(l+1)}=b({\bf X}^{(l)}),\;l=0,1,\cdots.
\end{equation}
with the coefficient matrix $D$ and the right term $b({\bf
X}^{(l)})$: $
\begin{pmatrix}
D,&b\end{pmatrix}=$
$$
\left(\begin{array}{ccccl}
\frac{h}{2}A&0&0&-B&\quad -\frac{h}{2}A{\bf u}^j+B\Phi^j+{\bf c}\\
-\delta B&\frac{h}{2}A&0&0&\quad \delta B{\bf u}^j-\frac{h}{2}A{\bf v}^j\\
0&-\delta r B&\frac{\tau}{2}A&-\frac{1}{2}A&\quad \delta r B {\bf v}^j-\frac{\tau}{2}
A{\bf w}^j-\frac{1}{2} A\Phi^j+2\tau {\bf V}({\bf u}^j,{\bf u}^{(l)})\\
\frac{1}{2}A&0&rB&0,&\quad\frac{1}{2}A{\bf u}^j-rB{\bf w}^j
\end{array}
\right)
$$
The initial guess is generally chosen as the value of the previous time step, i.e.
$${\bf X}^{(0)}=({\bf u}^{j}, {\bf v}^{j}, {\bf w}^{j}, {\bf\phi} ^{j})^{T}.$$

Unfortunately, this iteration is not convergent because the
coefficient matrix is degenerated. Note that $B$ is singular, and
its rank is $n-1$.  $A$ is a nonsingular matrix only if  $n$ is an
odd number. To take the row operation of the coefficient matrix
$D$, we suppose $n$ be an odd number and  $A^{-1}$ exist. Then
there is a permutation matrix $P$, so that
$$PD=R,$$
where $P$ is the unit lower triangular matrix, $R$ is an upper
triangular matrix.

The left multiplying (\ref{3.4}) by the permutation matrix $P$
yields an equivalent system
\begin{equation}\label{3.5}
\widetilde{D}{\bf X}^{(l+1)}=\widetilde{b}({\bf
X}^{(l)}),\;l=0,1,\cdots,
\end{equation}
where the augmented coefficient matrix now becomes $
(\widetilde{D},\widetilde{b})=P\cdot(D,b)= $
\begin{equation}\label{3.6}
\left(\begin{array}{cccll}
\frac{h}{2}A&0&0&-B&-\frac{h}{2}A{\bf u}^j+B\Phi^j+{\bf c}\\
0&\frac{h}{2}A&0&-\frac{2\delta}{h}BA^{-1}B&
\frac{2\delta}{h}BA^{-1}(B\Phi^j+{\bf c})
-\frac{h}{2}A{\bf v}^j\\
0&0&\frac{\tau A}{2}&-\frac{A}{2}-r(\frac{2\delta}{h}BA^{-1})^2B&r(\frac{2\delta}{h}B
A^{-1})^2(B\Phi^j+{\bf c})-\frac{\tau}{2}A{\bf w}^j-\frac{A}{2}\Phi^j +2\tau {\bf V}\\
0&0&0&\frac{2B}{h}+\frac{8\delta^2r}{h^3}(BA^{-1})^3B,& A{\bf
u}^j- \frac{8\delta^2r}{h^3}(BA^{-1})^3(B\Phi^j+{\bf c})
-4rBA^{-1}{\bf V}-\frac{\bf c}{h}
\end{array}
\right)
\end{equation}

{\bf Proposition 3.1}. The coefficient matrix $D$ of the iteration
(\ref{3.4}) is rank 1 deficient and
the deficiency  is relative to the potential function $\Phi$.\\
{\it Proof} \quad By (\ref{3.6})
\begin{align*}
\text{rank}(D)=&\text{rank}(\widetilde{D})\\
                =&3\text{rank}(A)+\text{rank}
([I_{n\times n}+4\delta ^2\frac{r}{h^2}(BA^{-1})^3]B)
\end{align*}

Note that $A$ is a full rank matrix and $B$ is a rank 1 deficient
matrix, we may take a proper $r$ ratio, so that
$$\text{det}(I_{n\times n}+4\delta ^2\frac{r}{h^2}(BA^{-1})^3)\ne 0,$$
$$\text{rank}([I_{n\times n}+4\delta ^2\frac{r}{h^2}(BA^{-1})^3]B) =\text{rank}(B).$$

Thus  $D$ is a rank 1 deficient matrix and it is obvious by the
(\ref{3.5}) that the deficiency is relative to the potential
function $\Phi$.

This proposition implies that we need one and only one more
condition to fix up the system (\ref{3.4}) and the condition is
related to the variable $\Phi$.  For example,
\begin{equation}\label{3.7}
\phi^{j+1}_i= a\; constant\quad \mbox{some a}\; i\in [1\; n].
\end{equation}
If we add this condition to the (\ref{3.3}), the rank of the
coefficient matrix $D$ is just full and the iteration is
convergent. It will be clarified by the following discussion and
the numerical experiments in section 5.

What we concern about is the solution of the KdV equation
(\ref{2.1}), i.e. the variable ${\bf u}^{j+1}$ in system
(\ref{3.3}). We hope that the added condition (\ref{3.7}) will not
change the value of the variable ${\bf u}^{j+1}$ in system
(\ref{3.3}),  which will be proved to be true in the following
proposition.

{\bf Proposition 3.2}.
The variables ${\bf u}^{j+1}$, ${\bf v}^{j+1}$, and $B\Phi^{j+1}$
in system (\ref{3.3}) are independent on the added condition
(\ref{3.7}), i.e.,  the value of $ \phi^{j+1}_i$.
They are only dependent on $B\Phi^{j}$, ${\bf u}^{j}$, ${\bf v}^{j}$.\\
{\it Proof.}\quad The four  iterative equations in the system
(\ref{3.5}) may be written as
\begin {align}
&\frac{h}{2}A {\bf u}^{(l+1)}-B\Phi^{(l+1)}
=-\frac{h}{2}A {\bf u}^{j}+B\Phi^{j}+{\bf c},\\
&\frac{h}{2}A {\bf v}^{(l+1)}-\frac{2\delta}{h}BA^{-1}B\Phi^{(l+1)}=
-\frac{h}{2}A {\bf v}^{j}+\frac{2\delta}{h}BA^{-1}(B\Phi^{j}+{\bf c}),\\
&\frac{\tau}{2}A {\bf w}^{(l+1)}-(\frac{A}{2}+r(\frac{2\delta}{h}BA^{-1})^2B)
\Phi^{(l+1)}=r(\frac{2\delta}{h}B
A^{-1 })^2(B\Phi^j+{\bf c})\notag\\
&\qquad\qquad \qquad-\frac{\tau}{2}A{\bf w}^j-\frac{A}{2}\Phi^j+
2\tau {\bf V}({\bf u}^{(l)},{\bf u}^j),\\
&(\frac{2}{h}I_{n\times n}+\frac{8\delta^2r}{h^3}(BA^{-1})^3)B\Phi^{(l+1)}=A{\bf u}^j-
\frac{8\delta^2r}{h^3}(BA^{-1})^3(B\Phi^j+{\bf c}) \notag\\
&\qquad\qquad\qquad-4rBA^{-1}{\bf V}({\bf u}^j,{\bf
u}^{(l)})-\frac{\bf c}{h}.
\end{align}
We now check the process of the iteration. Recall that the values
of ${\bf u}^{j}$, ${\bf v}^{j}$, and $B\Phi^{j}$ is given  and
${\bf u}^{(0)}={\bf u}^j$. By (3.11), $B\Phi^{(1)}$ is determined.
It is independent upon the value of $\phi^{j+1}_i$. The determined
$B\Phi^{(1)}$ and (3.8) fix up $A {\bf u}^{(1)}$. Because $A$ is
invertible, ${\bf u}^{(1)}$ is fixed up. For the same reason,
${\bf v}^{(1)}$ is also determined by (3.9). They are all
independent upon the  value of $\phi^{j+1}_i$. Substituting ${\bf
u}^{(1)}$ into (3.11) yields independence of $B\Phi^{(2)}$. As the
iteration goes on, we get three sequences ${\bf u}^{(l+1)}$, ${\bf
v}^{(l+1)}$, and $B\Phi^{(l+1)}$, $l=0, 1,\cdots,$ which are all
independent on the value of $ \phi^{j+1}_i$. Thus the convergent
point also has this property, namely, ${\bf u}^{j+1}$, ${\bf
v}^{j+1}$, and $B\Phi^{j+1}$ are independent on the value of $
\phi^{j+1}_i$.

This proposition assures the reliability of adding the condition
(\ref{3.7}) to the system (\ref{3.3}). Actually, the values of  $
u$, $v$ and $B\Phi$ are fixed up by the Preissman scheme, they
have nothing to do with the added condition. When to compute the
value of the variable $\Phi$, we need a  condition like
(\ref{3.7}) because the rank of the matrix $B$ is $n-1$.
 For convenience,  we may
take the condition (\ref{3.7}) as  a  boundary condition in
practical computation. That is  $\phi^{j+1}_1=0\; (or\;
\phi^{j+1}_{n+1}=0) $. The periodic numerical boundary condition
(\ref{3.2}) now becomes
\begin{equation}\label{3.12}
u_1^{j+1}=u_{n+1}^{j+1},\;\; v_1^{j+1}=v_{n+1}^{j+1}, \;\;
w_1^{j+1}=w_{n+1}^{j+1}, \;\; \phi_1^{j+1}=\phi_{n+1}^{j+1}+C=0,
\end{equation}
which makes the direct numerical implementation of the Preissman
scheme practical without changing the numerical solution of the
KdV equation.  The corresponding numerical results on soliton
collisions will be presented in the next section. {\bf Remark:}
\begin{itemize}
\item {The above analysis  is based on the condition that the number
of spatial grid points $n$ is an odd number. How to deal with the
case with an even number? In reference \cite{mars}, the numerical
experiments presented by Marsden et. al. also imply the question:
why sometimes the numerical results supported on odd spatial grid
points are quite different from that on even grid points ?}
\item{By (\ref{3.6}), we know that the variable  ${\bf w}^{j+1}$ will change if we
change the value of $\phi^{j+1}_i$ in the added condition
(\ref{3.7}). This means that the variable $w=\frac{1}{2}\phi_t
+\delta v_x +V^{'}(u)$ in the Preissman scheme is not fixed up.
Does the variable $w$ have the important physical meaning? If it
does, how to fix it up? }
\end{itemize}
 \sect{Some new  multisymplectic
schemes for the KdV equation}

In this section, we present several  new multisymplectic schemes
for the KdV equation.

Inspired by (\ref{3.6}), we obtain a new scheme for the KdV
equation (2.1)
\begin{equation}\label{4.1}
\begin{split}
{\bf p}^{j+1}&=M_1({\bf q}^j-\frac{1}{h}{\bf c})+M_2({\bf p}^j+{\bf c})+M_3(\frac{{\bf q}^j+{\bf q}^{j+1}}{4})^2,\\
{\bf q}^{j+1}&=-{\bf q}^{j}+\frac{2}{h}({\bf p }^j+{\bf
p}^{j+1}+{\bf c}).
\end{split}
\end{equation}
where ${\bf p}=B\Phi$, ${\bf q}=A{\bf u}$, $M_1\;M_2$, $M_3$ are
three constant matrixs, $M_1=\left[\frac{2}{h}I_{n}+
\frac{8\delta^2r} {h^3} (BA^{-1})^3\right]^{-1}$,
$M_2=-\frac{4\delta^2\tau}{h^3}(BA^{-1})^3-I_{n}$, $M_3=-2\eta
rM_1BA^{-1}$,
 ${\bf q}^2=(q_1^2, q_2^2, \cdots, q_n^2)^T$.

This scheme is equivalent to the multisymplectic Preissman scheme,
so it is also a multisymplectic scheme  and has an excellent
stability. Actually, it is composed of the first and the forth
line in (\ref{3.6}).  This scheme is more efficient than the
Preissman scheme because we need not to compute the variable $w$,
$v$ and $\phi$. we may  take  ${\bf p}=B\Phi$ and ${\bf q}=A{\bf
u}$ as new variables, furthermore  the condition (\ref{3.7}) does
not need. It enhances the conclusion that ${\bf u}^{j+1}$ is
independent on the additional numerical boundary value of $\phi$.
Moreover, Scheme (~\ref{4.1}) implies a natural iterative form
\begin{equation*}
\begin{split}
{\bf p}^{(l+1)}&=M_1{\bf q}^j+M_2({\bf p}^j+{\bf c})+M_3(\frac{{\bf q}^j+{\bf q}^{(l)}}{4})^2-M_1\frac{\bf c}{h},\\
{\bf q}^{(l+1)}&=-{\bf q}^{j}+\frac{2}{h}({\bf p }^j+{\bf
p}^{(l+1)}+{\bf c}).
\end{split}
\end{equation*}
 The computations of the iteration only involves  multiplication of matrices
and vectors, It avoids from solving the algebra equations which is
the main part of computation in other general implicit scheme such
as the Preissman scheme. After the convergent point $({\bf
p}^{k+1},{\bf q}^{k+1})$ is obtained, solving the equations $A{\bf
u}^{k+1}={\bf q}^{k+1}$ yields the numerical solution  at the
$k+1$th time step of the KdV equation (\ref{2.1}). If we want to
solve the system $B\Phi^{k+1}={\bf p}^{k+1}$ to get the numerical
results of the potential $\Phi^{k+1}$, the additional  condition
of $\phi$ like (\ref{3.7})  is also needed, for the coefficient
matrix $B$ is rank 1 deficient.

Eliminating the variable ${\bf p}$ in the scheme (~\ref{4.1}), we
have
\begin{align}\label{4.2}
\begin{split}
\frac{h}{2}({\bf q}^{j+1}+{\bf q}^j)=&(M_1+\frac{h}{2} M_2)({\bf
q}^j+{\bf q}^{j-1})\\
&+ M_3\left[(\frac{{\bf q}^{j+1}+{\bf q}^{j}}{4})^2+(\frac{{\bf
q}^{j}+{\bf q}^{j-1}}{4})^2 \right]+(I+M_2-\frac{1}{h}M_1){\bf c}.
\end{split}
\end{align}
Set ${\bf z}^{j+1}={\bf q}^{j+1}+{\bf q}^j$, then
\begin{equation}\label{4.3}
\frac{h}{2}{\bf z}^{j+1}=(M_1+\frac{h}{2}M_2){\bf z}^{j}+
M_3\left[(\frac{{\bf z}^{j+1}}{4})^2+(\frac{{\bf
z}^{j}}{4})^2\right] +(I+M_2-\frac{1}{h}M_1){\bf c}.
\end{equation}
Here the matrixs $M_1$, $M_2$ and $M_3$ are defined in
(\ref{4.1}).

This scheme is equivalent to the multisymplectic scheme
(\ref{4.1}). It contains only the variable $u$, thus it can be
viewed as a multisymplectic scheme for the original KdV equation
(\ref{2.1}). Its stability and capacity of long-time simulation
are the same with the multisymplectic Preissman scheme and scheme
(\ref{4.1}). Furthermore this scheme has more benefits such as
simple form, to practice easily and less computations.

It is worth mention that even if we modify the scheme (\ref{4.1})
into a real explicit scheme which don't need iterations when being
applied, the resulting scheme still have very nice numerical
performance which will be shown in next section. The explicit
scheme is
\begin{equation}\label{4.4}
\begin{split}
{\bf p}^{j+1}&=M_1{\bf q}^j+M_2{\bf p}^j+M_3(\frac{{\bf q}^j}{2})^2+(M_2-\frac{1}{h}M_1){\bf c},\\
{\bf q}^{j+1}&=-{\bf q}^{j}+\frac{2}{h}({\bf p }^j+{\bf p}^{j+1}+
{\bf c} ).
\end{split}
\end{equation}
Eliminating the variable ${\bf p}$ yields an explicit scheme for
the original KdV equation (\ref{2.1})
\begin{align}\label{4.5}
\begin{split}
\frac{h}{2}({\bf q}^{j+1}+{\bf q}^j)=&(M_1+\frac{h}{2}M_2)({\bf
q}^j+{\bf q}^{j-1})\\
&+ M_3\left[(\frac{{\bf q}^{j}}{2})^2+(\frac{{\bf q}^{j-1}}{2})^2
\right]+(I+M_2-\frac{1}{h}M_1){\bf c}.
\end{split}
\end{align}

If we modify the implicit term in scheme (\ref{4.2}) into an
explicit one, we obtain another explicit scheme for the KdV
equation
\begin{align}\label{4.6}
\begin{split}
\frac{h}{2}({\bf q}^{j+1}+{\bf q}^j)=&( M_1+\frac{h}{2}M_2)({\bf
q}^j+{\bf q}^{j-1})\\
&+ M_3\left[(\frac{{\bf q}^j}{2})^2+(\frac{{\bf q}^{j}+{\bf
q}^{j-1}}{4})^2 \right]+(I+M_2-\frac{1}{h}M_1){\bf c}.
\end{split}
\end{align}
But numerical results show that this is an unstable scheme.

All the schemes above are invalid provided the number $n$ of the
spatial grid points is even.  Next we introduce another method to
eliminate the auxiliary variables of the multisymplectic Preissman
scheme to get  two multisymplectic schemes for the KdV equation.
Both schemes are  valid whether $n$ is odd or even.

Let us state the multisymplectic Preissman scheme for equation
(2.2) in the form
\begin{align}
\frac{1}{2\triangle
t}&(u_{i-1}^{j}+u_{i}^{j}-u_{i-1}^{j-1}-u_{i}^{j-1})+
 \frac{1}{\triangle x}(w_{i}^{j-1}+w_{i}^{j}-w_{i-1}^{j-1}-w_{i-1}^{j})=0,\label{4.7}\\
 \frac{\delta}{\triangle x}(&u_{i}^{j-1}+u_{i}^{j}-u_{i-1}^{j-1}-u_{i-1}^{j})
-\frac{1}{2}(v_{i-1}^{j-1}+v_{i-1}^{j}+v_{i}^{j-1}+v_{i}^{j})=0,\label{4.8}\\
\frac{1}{2}(&u_{i}^{j-1}+u_{i}^{j}+u_{i-1}^{j-1}+u_{i-1}^{j})-
 \frac{1}{\triangle x}(\varphi_{i}^{j-1}+\varphi_{i}^{j}-
                            \varphi_{i-1}^{j-1}-\varphi_{i-1}^{j})=0,\label{4.9}\\
\frac{1}{2}(&w_{i-1}^{j-1}+w_{i-1}^{j}+w_{i}^{j-1}+w_{i}^{j})-
 \frac{1}{2\triangle t}(\varphi_{i-1}^{j}+\varphi_{i}^{j}-
                            \varphi_{i-1}^{j-1}-\varphi_{i}^{j-1})\notag\\
    &-\frac{\delta}{\triangle x}(v_{i}^{j-1}+v_{i}^{j}-
                            v_{i-1}^{j-1}-v_{i-1}^{j})=
     2V^{\prime}\big(\frac{1}{4}(u_{i-1}^{j-1}+u_{i-1}^{j}+u_{i}^{j-1}+u_{i}^{j})\big).
\label{4.10}
\end{align}
Eliminating the variable $w$ by (\ref{4.7}) and (4.10), we obtain
\begin{align}
&\frac{\triangle x}{4\triangle t}[u_{i+1}^j-u_{i+1}^{j-1}+
2(u_{i}^j- u_{i}^{j-1})+
u_{i-1}^j-u_{i-1}^{j-1}]+\frac{1}{2\triangle
t}[\phi_{i+1}^j-\phi_{i+1}^{j-1}\notag\\
 &-(\phi_{i-1}^j-\phi_{i-1}^{j-1})] +\frac{\delta}{\triangle
x}[v_{i+1}^j+v_{i+1}^{j-1}-2(v_{i}^j+v_{i}^{j-1})+v_{i-1}^j+v_{i-1}^{j-1}]= \notag\\
&-2[V^{\prime}\big
(\frac{1}{4}(u_{i+1}^{j-1}+u_{i+1}^{j}+u_{i}^{j-1}+u_{i}^{j})\big
)-V^{\prime}\big
(\frac{1}{4}(u_{i-1}^{j-1}+u_{i-1}^{j}+u_{i}^{j-1}+u_{i}^{j})\big
)].\label{4.11}
\end{align}
In the same manner, we may eliminate the variable $v$ by combining
(\ref{4.11}) and (\ref{4.8}) to obtain
\begin{align}
&\frac{\triangle x}{4\triangle t}[(u_{i+1}^j-u_{i+1}^{j-1})+
3(u_{i}^j- u_{i}^{j-1})+3(
u_{i-1}^j-u_{i-1}^{j-1})+u_{i-2}^j-u_{i-2}^{j-1}]\notag\\
&+\frac{1}{2\triangle t}[(\phi_{i+1}^j-\phi_{i+1}^{j-1})+
(\phi_{i}^j-\phi_{i}^{j-1})
 -(\phi_{i-1}^j-\phi_{i-1}^{j-1})-(\phi_{i-2}^j-\phi_{i-2}^{j-1})]\notag \\
 & +\frac{2\delta^2}{\triangle x^2}[(u_{i+1}^j+u_{i+1}^{j-1}-
 3(u_{i}^j+u_{i}^{j-1})+3(u_{i-1}^j+u_{i-1}^{j-1})- (u_{i-2}^j+u_{i-2}^{j-1})] \notag\\
&=-2[V^{\prime}\big
(\frac{1}{4}(u_{i+1}^{j-1}+u_{i+1}^{j}+u_{i}^{j-1}+u_{i}^{j})\big
)-V^{\prime}\big
(\frac{1}{4}(u_{i-1}^{j-1}+u_{i-1}^{j}+u_{i-2}^{j-1}+u_{i-2}^{j})\big
)],\label{4.12}
\end{align}
which together with (\ref{4.9}) yields a new 8-points scheme in
only variable $u$, by eliminating the variable $\phi$,
\begin{align}
&\frac{1}{4\triangle t}[(u_{i+1}^j+3u_{i}^j+3u_{i-1}^j+u_{i-2}^j)-
(u_{i+1}^{j-1}+3u_{i}^{j-1}+3u_{i-1}^{j-1}+u_{i-2}^{j-1})]\notag\\
&+\frac{\delta ^2}{\triangle
x^3}[(u_{i+1}^j-3u_{i}^j+3u_{i-1}^j-u_{i-2}^j)+
(u_{i+1}^{j-1}-3u_{i}^{j-1}+3u_{i-1}^{j-1}-u_{i-2}^{j-1})]\notag\\
&+\frac{1}{\triangle x}[V^{\prime}\big
(\frac{1}{4}(u_{i+1}^{j-1}+u_{i+1}^{j}+u_{i}^{j-1}+u_{i}^{j})\big
)-V^{\prime}\big
(\frac{1}{4}(u_{i-1}^{j-1}+u_{i-1}^{j}+u_{i-2}^{j-1}+u_{i-2}^{j})\big
)]\notag\\
&=0.\label{4.13}
\end{align}

Converting the implicit term  in above scheme to a explicit one,
we obtain a new explicit scheme for the KdV equation whose
remarkable numerical property will be shown in next section.
\begin{align}
&\frac{1}{4\triangle t}[(u_{i+1}^j+3u_{i}^j+3u_{i-1}^j+u_{i-2}^j)-
(u_{i+1}^{j-1}+3u_{i}^{j-1}+3u_{i-1}^{j-1}+u_{i-2}^{j-1})]\notag\\
&+\frac{\delta ^2}{\triangle
x^3}[(u_{i+1}^j-3u_{i}^j+3u_{i-1}^j-u_{i-2}^j)+
(u_{i+1}^{j-1}-3u_{i}^{j-1}+3u_{i-1}^{j-1}-u_{i-2}^{j-1})]\notag\\
&+\frac{1}{\triangle x}[V^{\prime}\big
(\frac{1}{2}(u_{i+1}^{j-1}+u_{i}^{j-1})\big )-V^{\prime}\big
(\frac{1}{2}(u_{i-1}^{j-1}+u_{i-2}^{j-1})\big )]=0.\label{4.14}
\end{align}

In the appendix of this paper we present another process of
eliminating the auxiliary variables to derive a 12-points scheme
\begin{align}
&\frac{1}{16\triangle
t}(u_{i+1}^{j+1}-u_{i+1}^{j-1}+3u_{i}^{j+1}-3u_{i}^{j-1}
   +3u_{i-1}^{j+1}-3u_{i-1}^{j-1}+u_{i-2}^{j+1}-u_{i-2}^{j-1})\notag\\
&+\frac{\delta^2}{4\triangle x^{3}}(u_{i+1}^{j+1}-3u_{i}^{j+1}
+3u_{i-1}^{j+1}-u_{i-2}^{j+1}+2u_{i+1}^{j}-6u_{i}^{j}\notag\\
&+6u_{i-1}^{j}
-2u_{i-2}^{j}+u_{i+1}^{j-1}-3u_{i}^{j-1}+3u_{i-1}^{j-1}-u_{i-2}^{j-1})\notag\\
+&\frac{1}{4\triangle x} \big[ V^{\prime}\big
(\frac{1}{4}(u_{i}^{j}+u_{i}^{j+1}+u_{i+1}^{j}+u_{i+1}^{j+1})\big
) - V^{\prime}\big
(\frac{1}{4}(u_{i-2}^{j}+u_{i-2}^{j+1}+u_{i-1}^{j}+u_{i-1}^{j+1})\big
)
              \notag\\
&+V^{\prime}\big
(\frac{1}{4}(u_{i}^{j-1}+u_{i}^{j}+u_{i+1}^{j-1}+u_{i+1}^{j})\big
) - V^{\prime}\big
(\frac{1}{4}(u_{i-2}^{j-1}+u_{i-2}^{j}+u_{i-1}^{j-1}+u_{i-1}^{j})\big
)\big]
\notag\\
&=0. \label{4.15}
\end{align}
Both schemes (\ref{4.13}) and (\ref{4.15}) are derived from the
the Preissman scheme (\ref{2.5}), thus they should be equivalent
to each other. Actually they can be derived from each other.

{\bf Remark}: The method introduced above can be applied to the
Preissman scheme, as well as other multisymplectic scheme, for
other Hamiltonian PEDs to obtain new schemes. For example, we can
get a nine-point scheme for the sine-Gordon equation, a six-point
scheme for the Schr\"{o}dinger equation, a forty-five-point for
the Kadomtsev-Petviashvili equation and so on. All these schemes,
except for the nine-point scheme, which was discussed by Marsden
et. al. in \cite{mars},  are new and expected to have excellent
numerical stability and capacity of long-time simulation.

\sect{Concluding Remarks} We analyze the multisymplectic scheme
for the KdV equation and find that the unconvergence of the
widely-used iteration method  to solve the resulting nonlinear
algebra system is due to the introduced potential function $\phi$.
We add a artificial numerical boundary condition on the original
periodic numerical boundary condition. It leads to  a new
numerical boundary condition (\ref{3.12}) which makes the
implementation of the Preissman scheme practical without changing
the numerical solution of the KdV equation. The numerical results
obtained with the presented numerical boundary condition show the
correctness of the condition and the merits of the multisymplectic
schemes. This method for analysis can be easily generalized to
other multisymplectic schemes and other PDEs.

 We obtain two new multisymplectic schemes for the KdV
equation, which are equivalent to, but more efficient than the
Preissman scheme.  We also develop a method to eliminate the
auxiliary variables of the Preissman scheme and get two equivalent
multisymplectic  schemes in only $u$ for the KdV equation. One is
a 12-point scheme and the other is an 8-point scheme.
 Compared with the
Zabusky-Kruskal scheme and the spectral method, the new
multisymplectic schemes are used to
 simulate the solitary waves.
Numerical results show that the multisymplecticity do bring the
finite differential schemes some benefits such as stability,
capacity for long time computation, and ability to preserve the
conservational laws.

At last we like point out the explicit schemes (\ref{4.5}) and
(\ref{4.14}) is also an excellent schemes for the KdV equation.
they can give the most accurate waveforms, which catch well up
with those in Figure 6. Furthermore, its the stability,  capacity
for long time computation and efficiency are much better than that
of the Zabusky-Kruskal scheme, as presented in subsection 5.5. We
are currently analyzing theoretically the stability, conservation
and other proprieties of the  explicit schemes.

{\it Remark}: During the preparation of this paper, Prof. R.
MacLachan has also derived the same results on the eight-points
for the KdV equation. We thank him for the discussions and many
important suggestions.

\newpage
\appendix{\bf Appendix}
\sect{The detail process of produce the 12-point scheme for the
KdV equation}
Let us state the multisymplectic Preissman scheme
for equation (2.2) in the form
\begin{align}
\frac{1}{2\triangle
t}&(u_{i-1}^{j}+u_{i}^{j}-u_{i-1}^{j-1}-u_{i}^{j-1})+
 \frac{1}{\triangle x}(w_{i}^{j-1}+w_{i}^{j}-w_{i-1}^{j-1}-w_{i-1}^{j})=0,\label{1}\\
 \frac{\delta}{\triangle x}(&u_{i}^{j-1}+u_{i}^{j}-u_{i-1}^{j-1}-u_{i-1}^{j})
-\frac{1}{2}(v_{i-1}^{j-1}+v_{i-1}^{j}+v_{i}^{j-1}+v_{i}^{j})=0,\label{2}\\
\frac{1}{2}(&u_{i}^{j-1}+u_{i}^{j}+u_{i-1}^{j-1}+u_{i-1}^{j})-
 \frac{1}{\triangle x}(\varphi_{i}^{j-1}+\varphi_{i}^{j}-
                            \varphi_{i-1}^{j-1}-\varphi_{i-1}^{j})=0,\label{3}\\
\frac{1}{2}(&w_{i-1}^{j-1}+w_{i-1}^{j}+w_{i}^{j-1}+w_{i}^{j})-
 \frac{1}{2\triangle t}(\varphi_{i-1}^{j}+\varphi_{i}^{j}-
                            \varphi_{i-1}^{j-1}-\varphi_{i}^{j-1})\notag\\
    &-\frac{\delta}{\triangle x}(v_{i}^{j-1}+v_{i}^{j}-
                            v_{i-1}^{j-1}-v_{i-1}^{j})=
     2V^{\prime}\big(\frac{1}{4}(u_{i-1}^{j-1}+u_{i-1}^{j}+u_{i}^{j-1}+u_{i}^{j})\big).
\label{4}
\end{align}
Taking $i=i+1$ in (\ref{4}), we obtain
\begin{align}
\frac{1}{2}(w_{i}^{j-1}+&w_{i}^{j}+w_{i+1}^{j-1}+w_{i+1}^{j})-
 \frac{1}{2\triangle t}(\varphi_{i}^{j}+\varphi_{i+1}^{j}-
                            \varphi_{i}^{j-1}-\varphi_{i+1}^{j-1})\notag\\
    &-\frac{\delta}{\triangle x}(v_{i+1}^{j-1}+v_{i+1}^{j}-
                            v_{i}^{j-1}-v_{i}^{j})=
     2V^{\prime}\big (\frac{1}{4}(u_{i}^{j-1}+u_{i}^{j}+u_{i+1}^{j-1}+u_{i+1}^{j})\big).
\label{5}
\end{align}
$\frac{(\ref{5})-(\ref{4})}{\triangle x}$ yields
\begin{align}
\frac{1}{2\triangle
x}&(w_{i+1}^{j-1}+w_{i+1}^{j}-w_{i-1}^{j-1}-w_{i-1}^{j})-
 \frac{1}{2\triangle t\triangle x}(\varphi_{i+1}^{j}-\varphi_{i+1}^{j-1}-
                            \varphi_{i-1}^{j}+\varphi_{i-1}^{j-1})\notag\\
    -&\frac{\delta}{\triangle x^{2}}(v_{i+1}^{j-1}+v_{i+1}^{j}-
                2v_{i}^{j-1}-2v_{i}^{j}+v_{i-1}^{j-1}+v_{i-1}^{j})=\notag\\
     &\frac{2}{\triangle x}
      \Big (V^{\prime}\big (\frac{1}{4}(u_{i}^{j-1}+u_{i}^{j}+u_{i+1}^{j-1}+u_{i+1}^{j})\big )
   - V^{\prime}\big (\frac{1}{4}(u_{i-1}^{j-1}+u_{i-1}^{j}+u_{i}^{j-1}+u_{i}^{j})\big )\Big
   ).
\label{6}
\end{align}
Let $i=i-1$ in (\ref{6}), then
\begin{align}\label{7}
\begin{split}
\frac{1}{2\triangle
x}&(w_{i}^{j-1}+w_{i}^{j}-w_{i-2}^{j-1}-w_{i-2}^{j})-
 \frac{1}{2\triangle t\triangle x}(\varphi_{i}^{j}-\varphi_{i}^{j-1}-
                            \varphi_{i-2}^{j}+\varphi_{i-2}^{j-1})\\
    -&\frac{\delta}{\triangle x^{2}}(v_{i}^{j-1}+v_{i}^{j}-
2v_{i-1}^{j-1}-2v_{i-1}^{j}+v_{i-2}^{j-1}+v_{i-2}^{j})\\
=&\frac{2}{\triangle x}
      \Big(V^{\prime}\big (\frac{1}{4}(u_{i-1}^{j-1}+u_{i-1}^{j}+u_{i}^{j-1}+u_{i}^{j})\big )
 - V^{\prime}\big (\frac{1}{4}(u_{i-2}^{j-1}+u_{i-2}^{j}+u_{i-1}^{j-1}+u_{i-1}^{j})\big )\Big
 ).
\end{split}
\end{align}
$\frac{(\ref{6})+(\ref{7})}{2}$ yields
\begin{align}
\frac{1}{4\triangle
x}&(w_{i+1}^{j-1}+w_{i+1}^{j}-w_{i-1}^{j-1}-w_{i-1}^{j}
   +w_{i}^{j-1}+w_{i}^{j}-w_{i-2}^{j-1}-w_{i-2}^{j})\notag\\
- &\frac{1}{4\triangle t\triangle
x}(\varphi_{i+1}^{j}-\varphi_{i+1}^{j-1}
-\varphi_{i-1}^{j}+\varphi_{i-1}^{j-1}+
       \varphi_{i}^{j}-\varphi_{i}^{j-1}-
         \varphi_{i-2}^{j}+\varphi_{i-2}^{j-1})\notag\\
    &-\frac{\delta}{2\triangle x^{2}}(v_{i+1}^{j-1}+v_{i+1}^{j}
-v_{i}^{j-1}-v_{i}^{j}-v_{i-1}^{j}-v_{i-1}^{j-1}
             + v_{i-2}^{j-1}+v_{i-2}^{j})\notag\\
=\frac{1}{\triangle x}& \Big (V^{\prime}\big
(\frac{1}{4}(u_{i}^{j-1}+u_{i}^{j}+u_{i+1}^{j-1}+u_{i+1}^{j})\big)
- V^{\prime}\big
(\frac{1}{4}(u_{i-2}^{j-1}+u_{i-2}^{j}+u_{i-1}^{j-1}+u_{i-1}^{j})\big
)\Big ). \label{8}
\end{align}
and, similiarly, taking $i=i+1$ in (\ref{2}), we have
\begin{equation}
 \frac{\delta}{\triangle x}(u_{i+1}^{j-1}+u_{i+1}^{j}-u_{i}^{j-1}-u_{i}^{j})-
\frac{1}{2}(v_{i}^{j-1}+v_{i}^{j}+v_{i+1}^{j-1}+v_{i+1}^{j})=0.\label{9}
\end{equation}
calculation of $\frac{(\ref{9})-(\ref{2})}{\triangle x}$ leads to
\begin{align}
 \frac{\delta}{\triangle x^{2}}&(u_{i+1}^{j-1}+u_{i+1}^{j}-2u_{i}^{j}-2u_{i}^{j-1}
 +u_{i-1}^{j-1}+u_{i-1}^{j})\notag\\
 &-\frac{1}{2\triangle x}(v_{i+1}^{j-1}+v_{i+1}^{j}-v_{i-1}^{j-1}-v_{i-1}^{j})=0.
\label{10}
\end{align}
which implies ($i=i-1$)
\begin{align}
 \frac{\delta}{\triangle x^{2}}&(u_{i}^{j-1}+u_{i}^{j}-2u_{i-1}^{j}-2u_{i-1}^{j-1}
 +u_{i-2}^{j-1}+u_{i-2}^{j})\notag\\
 &-\frac{1}{2\triangle x}(v_{i}^{j-1}+v_{i}^{j}-v_{i-2}^{j-1}-v_{i-2}^{j})=0.
\label{11}
\end{align}
$\frac{(\ref{10})-(\ref{11})}{\triangle x}$ yields
\begin{align}
 \frac{\delta}{\triangle x^{3}}&(u_{i+1}^{j-1}+u_{i+1}^{j}-3u_{i}^{j}-3u_{i}^{j-1}
 +3u_{i-1}^{j-1}+3u_{i-1}^{j}-u_{i-2}^{j}-u_{i-2}^{j-1})\notag\\
 &-\frac{1}{2\triangle x^{2}}(v_{i+1}^{j-1}+v_{i+1}^{j}-v_{i}^{j-1}-v_{i}^{j}
 -v_{i-1}^{j-1}-v_{i-1}^{j}+v_{i-2}^{j-1}+v_{i-2}^{j})=0.\label{12}
\end{align}

In the same manner, taking $i=i+1$ in (\ref{1}), we have
\begin{equation}
\frac{1}{2\triangle
t}(u_{i}^{j}+u_{i+1}^{j}-u_{i}^{j-1}-u_{i+1}^{j-1})+
 \frac{1}{\triangle x}(w_{i+1}^{j-1}+w_{i+1}^{j}-w_{i}^{j-1}-w_{i}^{j})=0.
\label{13}
\end{equation}
and the sum of the above formula and (1) is
\begin{align}
\frac{1}{2\triangle
t}&(u_{i+1}^{j}-u_{i+1}^{j-1}+2u_{i}^{j}-2u_{i}^{j-1}+
    u_{i-1}^{j}-u_{i-1}^{j-1})\notag\\
 &+\frac{1}{\triangle x}(w_{i+1}^{j-1}+w_{i+1}^{j}-w_{i-1}^{j-1}-w_{i-1}^{j})=0.
\label{14}
\end{align}

Combining (\ref{8}), (\ref{12}), (\ref{14}), we obtain
\begin{align}
-\frac{1}{8\triangle
t}&(u_{i+1}^{j}-u_{i+1}^{j-1}+3u_{i}^{j}-3u_{i}^{j-1}
   +3u_{i-1}^{j}-3u_{i-1}^{j-1}+u_{i-2}^{j}-u_{i-2}^{j-1})\notag\\
- &\frac{1}{4\triangle t\triangle
x}(\varphi_{i+1}^{j}-\varphi_{i+1}^{j-1}
-\varphi_{i-1}^{j}+\varphi_{i-1}^{j-1}+
       \varphi_{i}^{j}-\varphi_{i}^{j-1}-
         \varphi_{i-2}^{j}+\varphi_{i-2}^{j-1})\notag\\
    -&\frac{\delta ^2}{\triangle x^{3}}(u_{i+1}^{j-1}+u_{i+1}^{j}
-3u_{i}^{j}-3u_{i}^{j-1}+3u_{i-1}^{j-1}+3_{i-1}^{j}
             -u_{i-2}^{j}-u_{i-2}^{j-1})\notag\\
=&\frac{1}{\triangle x} \Big (V^{\prime}\big
(\frac{1}{4}(u_{i}^{j-1}+u_{i}^{j}+u_{i+1}^{j-1}+u_{i+1}^{j})\big
) - V^{\prime}\big
(\frac{1}{4}(u_{i-2}^{j-1}+u_{i-2}^{j}+u_{i-1}^{j-1}+u_{i-1}^{j})\big
)\Big ). \label{15}
\end{align}
This leads, if $j$ is replaced with $j+1$,
\begin{align}
-\frac{1}{8\triangle t} &(u_{i+1}^{j+1}-u_{i+1}^{j}+
3u_{i}^{j+1}-3u_{i}^{j}
   +3u_{i-1}^{j+1}-3u_{i-1}^{j}+u_{i-2}^{j+1}-u_{i-2}^{j})\notag\\
- &\frac{1}{4\triangle t\triangle
x}(\varphi_{i+1}^{j+1}-\varphi_{i+1}^{j}
-\varphi_{i-1}^{j+1}+\varphi_{i-1}^{j}+
       \varphi_{i}^{j+1}-\varphi_{i}^{j}-
         \varphi_{i-2}^{j+1}+\varphi_{i-2}^{j})\notag\\
    -&\frac{\delta^2}{\triangle x^{3}}(u_{i+1}^{j}+u_{i+1}^{j+1}
-3u_{i}^{j+1}-3u_{i}^{j}+3u_{i-1}^{j}+3_{i-1}^{j+1}
             -u_{i-2}^{j+1}-u_{i-2}^{j})\notag\\
=&\frac{1}{\triangle x} \Big (V^{\prime}\big
(\frac{1}{4}(u_{i}^{j}+u_{i}^{j+1}+u_{i+1}^{j}+u_{i+1}^{j+1})\big
) - V^{\prime}\big
(\frac{1}{4}(u_{i-2}^{j}+u_{i-2}^{j+1}+u_{i-1}^{j}+u_{i-1}^{j+1})\big
)\Big ). \label{16}
\end{align}
The sum of the above two formulas is
\begin{align}
-\frac{1}{8\triangle
t}&(u_{i+1}^{j+1}-u_{i+1}^{j-1}+3u_{i}^{j+1}-3u_{i}^{j-1}
   +3u_{i-1}^{j+1}-3u_{i-1}^{j-1}+u_{i-2}^{j+1}-u_{i-2}^{j-1})\notag\\
- &\frac{1}{4\triangle t\triangle
x}(\varphi_{i+1}^{j+1}-\varphi_{i+1}^{j-1}
+\varphi_{i}^{j+1}-\varphi_{i}^{j-1}-
       \varphi_{i-1}^{j+1}+\varphi_{i-1}^{j-1}+
         \varphi_{i-2}^{j-1}-\varphi_{i-2}^{j+1})\notag
\end{align}
\begin{align}
-&\frac{\delta^2}{\triangle x^{3}}(u_{i+1}^{j+1}+2u_{i+1}^{j}
+u_{i+1}^{j-1}-3u_{i}^{j+1}-6u_{i}^{j}-3u_{i}^{j-1}\notag\\
&+6u_{i-1}^{j}+3u_{i-1}^{j-1}
+3u_{i-1}^{j+1}-2u_{i-2}^{j}-u_{i-2}^{j+1}-u_{i-2}^{j-1})\notag\\
=&\frac{1}{\triangle x} \Big (V^{\prime}\big
(\frac{1}{4}(u_{i}^{j}+u_{i}^{j+1}+u_{i+1}^{j}+u_{i+1}^{j+1})\big
) - V^{\prime}\big
(\frac{1}{4}(u_{i-2}^{j}+u_{i-2}^{j+1}+u_{i-1}^{j}+u_{i-1}^{j+1})\big)
              \notag\\
&+V^{\prime}\big
(\frac{1}{4}(u_{i}^{j-1}+u_{i}^{j}+u_{i+1}^{j-1}+u_{i+1}^{j})\big
) - V^{\prime}\big
(\frac{1}{4}(u_{i-2}^{j-1}+u_{i-2}^{j}+u_{i-1}^{j-1}+u_{i-1}^{j})\big
)\Big ).\label{17}
\end{align}

Meanwhile, we take $j=j+1$ in (\ref{3})
\begin{equation}
\frac{1}{2}(u_{i}^{j}+u_{i}^{j+1}+u_{i-1}^{j}+u_{i-1}^{j+1})-
 \frac{1}{\triangle x}(\varphi_{i}^{j}+\varphi_{i}^{j+1}-
                            \varphi_{i-1}^{j}-\varphi_{i-1}^{j+1})=0.
\label{18}
\end{equation}
$\frac{(\ref{18})-(\ref{3})}{\triangle t}$ yields
\begin{equation}
\frac{1}{2\triangle
t}(u_{i}^{j+1}+u_{i-1}^{j+1}-u_{i}^{j-1}-u_{i-1}^{j-1})-
 \frac{1}{\triangle x\triangle t}(\varphi_{i}^{j+1}-\varphi_{i-1}^{j+1}-
             \varphi_{i}^{j-1}+\varphi_{i-1}^{j-1})=0.\label{19}
\end{equation}
Taking $i=i+1$ in (\ref{19}), we obtain
\begin{equation}
\frac{1}{2\triangle
t}(u_{i}^{j+1}+u_{i+1}^{j+1}-u_{i+1}^{j-1}-u_{i}^{j-1})-
 \frac{1}{\triangle x\triangle t}(\varphi_{i+1}^{j+1}-\varphi_{i}^{j+1}-
                  \varphi_{i+1}^{j-1}+\varphi_{i}^{j-1})=0.\label{20}
\end{equation}
By (\ref{19})+(\ref{20}), we have
\begin{align}
\frac{1}{2\triangle t}(u_{i+1}^{j+1}&+2u_{i}^{j+1}+u_{i-1}^{j+1}-2
          u_{i}^{j-1}-u_{i-1}^{j-1}-u_{i+1}^{j-1})\notag\\
 &-\frac{1}{\triangle x\triangle t}(\varphi_{i+1}^{j+1}-\varphi_{i+1}^{j-1}-
           \varphi_{i-1}^{j+1}+\varphi_{i-1}^{j-1})=0.\label{21}
\end{align}
which is
\begin{align}
\frac{1}{2\triangle t}(u_{i}^{j+1}&+2u_{i-1}^{j+1}+u_{i-2}^{j+1}-2
          u_{i-1}^{j-1}-u_{i-2}^{j-1}-u_{i}^{j-1})\notag\\
 &-\frac{1}{\triangle x\triangle t}(\varphi_{i}^{j+1}-\varphi_{i}^{j-1}-
                   \varphi_{i-2}^{j+1}+\varphi_{i-2}^{j-1})=0,\label{22}
\end{align}
if $i$ is replaced with$i-1$.

Combining (\ref{21}),(\ref{22}) and (\ref{17}), we obtain a new
multisymplectic twelve points scheme for the KdV equation (2.1)
\begin{align}
\frac{1}{16\triangle
t}&(u_{i+1}^{j+1}-u_{i+1}^{j-1}+3u_{i}^{j+1}-3u_{i}^{j-1}
   +3u_{i-1}^{j+1}-3u_{i-1}^{j-1}+u_{i-2}^{j+1}-u_{i-2}^{j-1})\notag\\
+&\frac{\delta^2}{4\triangle x^{3}}(u_{i+1}^{j+1}-3u_{i}^{j+1}
+3u_{i-1}^{j+1}-u_{i-2}^{j+1}+2u_{i+1}^{j}-6u_{i}^{j}\notag\\
&+6u_{i-1}^{j}
-2u_{i-2}^{j}+u_{i+1}^{j-1}-3u_{i}^{j-1}+3u_{i-1}^{j-1}-u_{i-2}^{j-1})\notag\\
+&\frac{1}{4\triangle x} \big[ V^{\prime}\big
(\frac{1}{4}(u_{i}^{j}+u_{i}^{j+1}+u_{i+1}^{j}+u_{i+1}^{j+1})\big
) - V^{\prime}\big
(\frac{1}{4}(u_{i-2}^{j}+u_{i-2}^{j+1}+u_{i-1}^{j}+u_{i-1}^{j+1})\big
)
              \notag\\
+&V^{\prime}\big
(\frac{1}{4}(u_{i}^{j-1}+u_{i}^{j}+u_{i+1}^{j-1}+u_{i+1}^{j})\big
) - V^{\prime}\big
(\frac{1}{4}(u_{i-2}^{j-1}+u_{i-2}^{j}+u_{i-1}^{j-1}+u_{i-1}^{j})\big
)\big]
\notag\\
=0&. \label{23}
\end{align}

\begin{thebibliography}{99}
\bibitem{zk} N.J.Zabusky \& M.D.Kruskal, Interaction of "soliton" in a Collisionless Plasma
            and Recurrence if Initial States, Phys. Rev. Letters, 15, 240-243, 1965.
\bibitem{fouri} S.B.Wineberg, J.F.Mcgrath, E.F.Gabl, ect., Implicit Spectral Methods for
            Wave Propagation Problems, Jour. Comp. Phys. 97, 311-336, 1991.
\bibitem{rw} R. Winther, A conservative finite element method for
             the Korteweg-de Vries equation, Math. Compu. 34, 23-43, 1980.
\bibitem{hai}  Ernst Hairer, Christian Lubich, Invariant tori of
            dissipatively perturbed Hamiltonian systems under symplectic
            discretization, Appl. Numer. Math. 29, 57-71,1999.
\bibitem{san} Sanz-Serna J M, Calvo M P. Numerical hamiltonian problem. Chapman and Hall, London, 1994
\bibitem{fq}  Feng K, Qin M Z, The symplectic methods for computation of Hamiltonian systems,
            In Zhu Y L, Guo Ben-Yu, ed, Proc Conf on Numerical Methods for PDEs,
            Berlin: Springer, 1987, 1-37, Lecture notes in Math, 1297.
\bibitem{mars} J.E.Marsden, G.P.Patrick and S.Shkoller, Multisymplectic geometry, variational
            integrators, and Nonlinear PDEs, Comm. Math. Phys, 199, 351-395(1998).
\bibitem{bs} T.J.Bridges, S.Reich, Multi-symplectic Integrators: numerical schemes for Hamiltonian
            PDEs that conserve symplecticity, Physics letter A, 2001, 284(4-5):184-193.
\bibitem{zhao} P.F. Zhao, M.Z. Qin, multisymplectic Geometry and
            Multisymplectic Preissman Scheme for the KdV Equation, J. Phys. A:
            Math. Gen, 33, 3613-1626, 2000.
\bibitem{jb} Jing-Bo Chen, New schemes for the nonlinear Schrodinger equation,  Applied Mathematics
            and Computation, 124(3), 371-379, 2001.
\bibitem{sr} S. Reich, Notes on Numerical Methods for Hamiltonian PDEs, Reading Material
                 of International Workshop on structure-Preserving Algorithms, Vol. 6(3),
                 153-174, 2001.
\bibitem{my} Yushun Wang, Mengzhao Qin, Multisymplectic Geometry and Multisymplectic
             Scheme for the Nonlinear Klein Gordon Equation. Journal of the
             Physical Society  of Japan.Vol.70, No.3, 653-661, March 2001.
\end{thebibliography}
\end{document}